
\magnification=\magstep1
\hsize=13cm
\vsize=20cm
\overfullrule 0pt
\baselineskip=13pt plus1pt minus1pt
\lineskip=3.5pt plus1pt minus1pt
\lineskiplimit=3.5pt
\parskip=4pt plus1pt minus4pt

\def\negenspace{\kern-1.1em}


\newcount\secno
\secno=0
\newcount\susecno
\newcount\fmno\def\z{\global\advance\fmno by 1 \the\secno.
                       \the\susecno.\the\fmno}
\def\section#1{\global\advance\secno by 1
                \susecno=0 \fmno=0
                \centerline{\bf \the\secno. #1}\par}
\def\subsection#1{\medbreak\global\advance\susecno by 1
                  \fmno=0
       \noindent{\the\secno.\the\susecno. {\it #1}}\noindent}
%
\def\sqr#1#2{{\vcenter{\hrule height.#2pt\hbox{\vrule width.#2pt
height#1pt \kern#1pt \vrule width.#2pt}\hrule height.#2pt}}}
\def\square{\mathchoice\sqr64\sqr64\sqr{4.2}3\sqr{3.0}3}
\setbox1=\hbox{$\square$}
\def\dalem{\raise 1pt\copy1}
%


\newcount\refno
\refno=1
\def\y{\the\refno}
\def\myfoot#1{\footnote{$^{(\y)}$}{#1}
                 \advance\refno by 1}


\def\newref{\vskip 1pc 
            \hangindent=2pc
            \hangafter=1
            \noindent}

\def\neq{\hbox{$\,$=\kern-6.5pt /$\,$}}






\newcount\secno
\secno=0
\newcount\fmno\def\z{\global\advance\fmno by 1 \the\secno.
                       \the\fmno}
\def\sectio#1{\medbreak\global\advance\secno by 1
                  \fmno=0
       \noindent{\the\secno. {\it #1}}\noindent}



\def\vta{\vartheta}

\def\a{\alpha}
\def\b{\beta}

\def\vta{\vartheta}

\null
\bigskip\bigskip\bigskip\bigskip
\centerline{\bf HYPERFLUID --- A MODEL OF} 
\centerline{\bf CLASSICAL MATTER WITH HYPERMOMENTUM}
\bigskip  
\centerline{by}  
\bigskip  
\centerline{Yuri N.\ Obukhov$^{*\diamond}$ and Romualdo Tresguerres}
\bigskip  
\bigskip  
\centerline{ Institute for Theoretical Physics, University of  
Cologne, D--50923 K\"oln}  
\centerline{Germany}

\bigskip\bigskip  
\bigskip  
\centerline{\bf Abstract} 

A variational theory of a continuous medium is developed the elements of which 
carry momentum and hypermomentum ({\it hyperfluid}). It is shown that the 
structure of the sources in metric-affine gravity is predetermined by the 
conservation identities and, when using the {\it Weyssenhoff ansatz}, these 
explicitly yield the hyperfluid currents. 

\bigskip\bigskip  
\bigskip  
\bigskip\bigskip  

\bigskip\bigskip  
\bigskip  
\bigskip\bigskip  

\vfill

\noindent $^{*})$ Permanent address: Department of Theoretical Physics, 
Moscow State University, 117234 Moscow, Russia.

\noindent $^{\diamond})$ Alexander von Humboldt Fellow.

\eject  

\sectio{\bf Introduction}
\bigskip
\bigskip

Fluid models (see, e.g., [1-3]) play an important role in gravitational 
theory, providing a convenient description of classical matter in terms 
of hydrodynamical notions. Various applications can be mentioned, starting 
>from cosmology and astrophysics and including approximation schemes of the
post-Newtonian formalism. 

Spin fluids [4-8], or continuous media with internal angular momentum, form 
the basis for understanding the physics of polarizable matter. This type of a 
classical (i.e., non-quantum) source occurs most naturally within the framework
of Poincar\'e gauge gravity [9,10].

A further generalization leads to metric-affine gravity based on the gauge 
theory for the general affine group $GA(4,R)$ [11-17]. This article presents 
an attempt to construct a variational theory of a {\it hyperfluid} --- a 
continuous medium the elements of which are characterized by a nontrivial 
hypermomentum density. 

\bigskip
\bigskip
\sectio{\bf Conservation identities and phenomenological approach to a 
hyperfluid}
\bigskip
\bigskip
The gravitational variables of metric-affine gravity are the
forms ($g_{\alpha\beta}$, $\vta^\alpha$, $\Gamma_\alpha{}^\beta$) with an 
appropriate transformation behavior under the local $GL(4,R)$ group. 
The metric $g_{\alpha\beta}$ is a 0-form, the coframe $\vartheta^\alpha$ and 
the connection $\Gamma_\alpha{}^\beta$ are 1-forms. The list of the field
strengths includes the {\it nonmetricity} 1-form $Q_{\alpha\beta}:=-
Dg_{\alpha\beta}$ besides the 2-forms of torsion $T^\alpha$ and curvature 
$R_\alpha{}^\beta$ [18, 19].

Let us start the discussion of the generalized fluid by displaying the
gravitational field equations. For the most general gravitational
Lagrangian 4-form $V=V$($g_{\alpha\beta}$, $\vta^{\alpha}$, 
$Q_{\alpha\beta}$, $T^{\alpha}$, $R_{\alpha}^{\ \beta}$) they read:
$$
2{\delta V \over \delta g_{\alpha\beta}}=-\sigma^{\alpha\beta},\eqno(\z)
$$
$$
{\delta V \over \delta \vta^{\alpha}}=-\Sigma_{\alpha},\eqno(\z)
$$
$$
{\delta V \over \delta \Gamma_{\alpha}^{\ \beta}}=-
\Delta^{\alpha}_{\ \beta},\eqno(\z)
$$
The left-hand sides of the field equations (2.1)-(2.3) are not quite
independent. They satisfy the {\it identities} which result from the
Noether theorem for general coordinate (diffeomorphism) and local
$GL(4,R)$ invariance of the gravitational action,
$$
D{\delta V \over \delta\vta^{\alpha}}\equiv 
(e_\alpha\rfloor T^\beta)\wedge{\delta V \over \delta\vta^{\beta}} +
(e_\alpha\rfloor R_{\beta}{}^{\gamma})\wedge
{\delta V \over \delta\Gamma_{\beta}^{\ \gamma}} -(e_{\alpha}\rfloor 
Q_{\beta\gamma}){\delta V \over \delta g_{\beta\gamma}},\eqno(\z)
$$
$$
D{\delta V \over \delta\Gamma_{\alpha}{}^{\beta}} + 
\vartheta^{\alpha}\wedge{\delta V \over \delta\vta^{\beta}} -  
2g_{\beta\gamma}{\delta V \over \delta g_{\alpha\gamma}}\equiv 0,\eqno(\z)
$$  
where the vectors $e_\alpha$ constitute the (anholonomic) {\it frame} 
($e_{\alpha}\rfloor\vartheta^{\beta}=\delta^{\beta}_{\alpha}$). 

The right-hand sides of the gravitational field equations are represented by
the {\it matter currents}: the metric stress-energy $\sigma^{\a\b}$, the 
canonical energy-momentum $\Sigma_\a$, and the hypermomentum $\Delta^\a{}_\b$,
the latter is asymmetric in $\alpha$ and $\beta$. 

In a self-consistent {\it variational} framework the matter currents should
arise quite generally from a matter Lagrangian $L$ as variational derivatives
$$
\sigma^{\alpha\beta} := 2{{\delta L}\over{\delta g_{\alpha\beta}}},\ \ \ \ 
\Sigma_{\alpha}:={{\delta L}\over{\delta\vartheta^{\alpha}}},\ \ \ \ 
\Delta^{\alpha}{}_{\beta}
:= {{\delta L}\over{\delta\Gamma_{\alpha}{}^{\beta}}} .\eqno(\z)
$$  

However, let us suppose that the precise form of the matter Lagrangian is 
unknown. To what extent can one determine the structure of the matter currents?
Such an approach (which could be called {\it phenomenological}) proved to be
useful in Einstein's general relativity theory, as well as in Poincar\'e gauge
gravity. We will demonstrate that, in fact, the conservation identities provide
quite a powerful tool for establishing the theory of a fluid with 
hypermomentum.

In the absence of an explicit matter Lagrangian $L$, the standard 
Noether framework is not available for discussing the symmetry properties of 
matter. However, now the identities (2.4)-(2.5) play a central role, giving 
the conditions which provide the mathematical self-consistency of the 
gravitational theory. Indeed, let us substitute (2.1)-(2.3) into these 
geometrical identities. As a result we obtain two {\it equations} for 
the matter currents,
$$
D\Sigma_\alpha=(e_\alpha\rfloor T^\beta)\wedge\Sigma_\beta+
  (e_\alpha\rfloor R_{\beta}{}^{\gamma})\wedge\Delta^{\beta}{}_{\gamma}
-{1\over2}(e_{\alpha}\rfloor  
Q_{\beta\gamma})\,\sigma^{\beta\gamma} ,\eqno(\z)
$$
and
$$
D\Delta^{\alpha}{}_{\beta} +  \vartheta^{\alpha}\wedge\Sigma_{\beta} -  
g_{\beta\gamma}\>\sigma^{\alpha\gamma}=0.\eqno(\z)
$$  

Unlike (2.4)-(2.5) these are not identically satisfied, but instead should 
be understood as constraints on the {\it matter variables} (unspecified as yet)
out of which the matter currents are constructed. 

As it is clear from (2.2), (2.3), the currents ($\Sigma_{\alpha}$,\ 
$\Delta^{\alpha}_{\ \beta}$) are associated to the {\it translational} and
{\it rotational-deformational} (local $GL(4,R)$) gravitational degrees of
freedom, respectively. Thus they represent the essential physical constituents 
of the theory, namely the $GA(4,R)$--matter currents. In contrast, the
current $\sigma^{\alpha\beta}$ is a secondary object, which is
proved by the fact that equation (2.8) determines the
current $\sigma^{\alpha\beta}$, provided $\Sigma_{\alpha}$ and  
$\Delta^{\alpha}_{\ \beta}$ are given.

Let us present subsequently the phenomenological description of the hyperfluid.
We will consider it as a continuous medium the elements of which are 
characterized by the density of the classical ``charge" of the relevant gauge 
group --- i.e., by the pair ($P_{\alpha}$, $J^{\alpha}_{\ \beta}$) in our case.
These quantities represent the 4-momentum and the hypermomentum of a fluid 
element, respectively. As usually in hydrodynamics, the 4-velocity vector 
field $u^\alpha$ is defined by the flow of the fluid. In the language of 
exterior calculus, it is more convenient to start with a {\it flow 3-form} 
$u$ [20]. The components of the velocity are then defined by 
$$
u_{\alpha}:=e_{\alpha}\rfloor\ast u,\eqno(\z)
$$
where $\ast$ is the Hodge dual operator. The 4-velocity is assumed to be a 
timelike vector with unit length, $u^{\alpha}u_{\alpha}=1$. This translates
into the condition
$$
\ast u\wedge u = \eta,\eqno(\z)
$$
where the volume 4-form is defined by means of the Hodge dual,
$\eta:=\ast 1$. The current of a $GA(4,R)$-charged fluid,
produced by the flow $u$, is simply a 3-form $u(P_{\alpha}$, 
$J^{\alpha}_{\ \beta})$. We will assume that this
{\it phenomenological} current describes the right-hand side of (2.2)-(2.3). 
Thus the hyperfluid matter current 3-forms are given by 
$$
\Sigma_{\alpha}=uP_{\alpha},\eqno(\z)
$$
$$
\Delta^{\alpha}_{\ \beta}=uJ^{\alpha}_{\ \beta}.\eqno(\z)
$$

The representation (2.11)-(2.12) (which can be called the {\it generalized
Weyssenhoff ansatz}) proved to be viable both in Einstein theory, and
in Poincar\'e gauge gravity. In the former case only the translations 
effectively form the gauge symmetry group of space-time. Hence the relevant 
fluid is characterized by the matter currents $\Sigma_{\ \alpha}=uP_{\alpha}$, 
$\Delta^{\alpha}_{\ \beta}=0$, which describe an ordinary structureless
continuous medium. In Poincar\'e gravity, the matter currents are 
$\Sigma_{\alpha}=uP_{\alpha}$, $\Delta^{\alpha}_{\ \beta}=
uS^{\alpha}_{\ \beta}$, with $S^{\alpha\beta}=-S^{\beta\alpha}$ (spin), and 
these describe the Weyssenhoff spin fluid [4,8,20]. The antisymmetric part of 
the hypermomentum charge in (2.12) represents the spin density, 
$J_{[\alpha\beta]}:=S_{\alpha\beta}$.

Remarkably, the conservation identities (2.7)-(2.8) contain much of the 
information necessary for establishing the hyperfluid theory. At first, let
us use (2.8) and find the structure of the canonical energy-momentum and 
the metrical stress of the hyperfluid. For this purpose we notice that the 
antisymmetric part of (2.8),
$$
g_{\gamma [\alpha }D(uJ^\gamma {}_{\beta ]}) +\vartheta_{[\alpha}\wedge 
uP_{\beta ]} =0\,,\eqno(\z)
$$
is easily solved with respect to the 4-momentum,
$$
P_{\alpha}=\ast(u\wedge\pi_{\alpha})\,,\qquad 
\pi_{\alpha}=\varepsilon\vartheta_{\alpha} + 
2\vartheta^{\beta}g_{\gamma [\alpha }\dot{J}^\gamma {}_{\beta ]} \,,\eqno(\z)
$$
where hereafter for any covariant object $\Phi^{\alpha\cdot\cdot\cdot}
_{\beta\cdot\cdot\cdot}$ the dot denotes $\dot{\Phi}^{\alpha\cdot\cdot\cdot}
_{\beta\cdot\cdot\cdot}:=-\ast D(u\Phi^{\alpha\cdot\cdot\cdot}
_{\beta\cdot\cdot\cdot})$, and the scalar $\varepsilon:=u^{\alpha}P_{\alpha}$ 
has the meaning of the rest energy density.

The substitution of (2.14) back into (2.11) yields the explicit form of the 
canonical energy-momentum current of the hyperfluid:
$$
\Sigma_{\alpha}=u\ast \Big[ u\wedge (\varepsilon\vartheta_{\alpha} +
2\vartheta^{\beta}g_{\gamma [\alpha }\dot{J}^\gamma {}_{\beta ]})\Big]
\,.\eqno(\z)
$$
In turn, the symmetric part of (2.8) yields the metric stress current
$$
\sigma_{\alpha\beta} = \eta \Big[\varepsilon u_\alpha u_\beta + 
u^\lambda u_{(\alpha}g_{\beta)\gamma}\dot{J}^{\gamma}{}_{\lambda} 
+ h_{\gamma (\alpha }\dot{J}^\gamma {}_{\beta )}\Big]\,,\eqno(\z)
$$
where $h^{\alpha}_{\beta}:=\delta^{\alpha}_{\beta} - u^\alpha u_\beta $ is the 
standard projector on the subspace orthogonal to the 4-velocity $u^\alpha$.

It is worthwhile to note that, when the nonmetricity is zero (i.e., in a
Riemann-Cartan space-time), only spin contributes to the canonical 
energy-momentum current (2.15). This becomes clear after rewriting $\pi_\alpha$
in (2.14) as $\pi_{\alpha}=\varepsilon\vartheta_{\alpha} + 
2\vartheta^{\beta}\dot{S}_{\alpha\beta} + 
2Q_{\gamma[\alpha}J^{\gamma}_{\ \beta]}\ast(u\wedge\vartheta^{\beta})$.

So far no non-gravitational interaction was assumed between the elements
of the fluid. Thus the model above describes the case of {\it incoherent} 
(or ``dust") matter with hypermomentum. The interparticle 
interactions manifest themselves in the additional {\it stress} 
term in the canonical energy-momentum tensor. Within the framework of 
the phenomenological approach under consideration, this additional term 
should be postulated separately. The simplest possibility is to adopt 
the {\it ideal} fluid postulate, which states that the stress produced by 
the interparticle interactions is given by the isotropic pressure $p$. 
Hence, the final canonical energy-momentum current of the {\it ideal 
hyperfluid} reads
$$\Sigma_{\alpha}=\, p\ast u\wedge (e_\alpha\rfloor u ) 
+u\ast \Big[ u\wedge (\varepsilon\vartheta_{\alpha} +
2\vartheta^{\beta}g_{\gamma [\alpha }\dot{J}^\gamma {}_{\beta ]})\Big]
\,.\eqno(\z)$$
The metric stress current (2.16) is modified then, according to (2.8), by 
including the stress term $-ph_{\alpha\beta}\eta$.

In the next section we present a self-consistent variational theory of
the hyperfluid, constructed along the lines of the early spin fluid models.

\bigskip
\bigskip
\sectio{\bf Variational principle for the ideal hyperfluid}
\bigskip
\bigskip

We will construct the model of a hyperfluid in generalizing the variational
theory of the Weyssenhoff spin fluid [21]. The motion of the medium will
be described, as usual, by its 4-velocity $u^{\alpha}$ and three vectors 
$b^{\alpha}_{A},\ A=1,2,3$, attached to each element of the fluid. However, 
unlike the {\it rigid} triad of the spin fluid model (which can only 
rotate), we will assume, in accordance with the $GA(4,R)$ gauge approach, 
that the material frame $b^{\alpha}_{A}$ is {\it elastic} --- in the sense 
that it can undergo arbitrary deformations during the motion of the fluid. 
This transition from a rigid to a deformable material frame is in fact well 
known in the elasticity theory as the transition from a Cosserat continuum 
[22] to the elastic medium {\it with micro-structure} of Mindlin [23]. 
Like the fluid velocity, we will describe the material frame by the 3-form
$b_A$, such that $b^{\alpha}_{A}:=\ast(b_{A}\wedge\vartheta^{\alpha})$, cf.
(2.9).

Wishing to preserve the Weyssenhoff model as a limiting case of the
theory under consideration, we will assume that the 4-velocity is normalized
and orthogonal to the material frame,
$$
\ast u\wedge u=\eta,\ \ \ \ast u\wedge b_{A}=0.\eqno(\z)
$$

Technically, it will be convenient also to introduce explicitly the dual
material triad --- the material co-frame $b^{A}_{\alpha},\ A=1,2,3$ which 
satisfies $b^{A}_{\alpha}b^{\alpha}_{B}=\delta^{A}_{B}.$ In the exterior
form language this will be described by the 1-form $b^A$ (so that 
$b^{A}_{\alpha}:=e_{\alpha}\rfloor b^{A}$) which is dual to the 3-form $b_A$,
$$
b^{B}\wedge b_{A} = \delta^{B}_{A}\eta .\eqno(\z)
$$

We will treat the pair $(b_{A}, b^{B})$ as independent dynamical
variables and (3.2) as the constraint, which simplifies greatly the non-trivial
problem of raising and lowering indices in the metric-affine approach.

The internal structure of the hyperfluid is characterised by the following
scalar variables: the particle density $\rho$, the specific entropy $s$, the 
Lin variable $X$ [24,25] (used to identify particles), and the specific 
hypermomentum density $\mu^{A}_{\ B}$ which is the direct generalisation of 
the spin density variable of the old Weyssenhoff model. As usually, we assume 
that the number of particles is not changed and that the entropy and the 
identity of particles is conserved during the motion of the fluid. These 
conditions are manifested in the form of constraints
$$
d(\rho u)=0,\eqno(\z)
$$
$$
u\wedge dX=0,\eqno(\z)
$$
$$
u\wedge ds=0,\eqno(\z)
$$
which will be introduced into the variational principle by means of the method
of the Lagrange multipliers. 

Now we are in a position to write the Lagrangian 4-form of the hyperfluid:
$$
L=\varepsilon(\rho, s, \mu^{A}_{\ B})\eta - 
{1\over 2}\rho\mu^{A}_{\ B}b^{B}_{\alpha}u\wedge Db^{\alpha}_{A} -
$$
$$
- \rho u\wedge d\lambda_{1} + \lambda_{2}u\wedge dX +
\lambda_{3}u\wedge ds + \lambda_{0}(\ast u\wedge u - \eta) +
$$
$$
+ \lambda^{A}(\ast u\wedge b_{A}) + \lambda^{A}_{B}(b^{B}\wedge b_{A} - 
\eta\delta^{B}_{A}) + \tilde{\lambda}_{A}(b^{A}\wedge u).\eqno(\z)
$$
The first line is most essential physically, representing the internal 
energy density $\varepsilon$ which is assumed to be the function of the
particle density, entropy and the specific hypermomentum density, and 
the kinetic energy (the second term in (3.6)), which represents in fact
the sum of well known rotational and elastic deformation energy terms.
The remaining terms in (3.6) describe constraints (the last term is necessary
to ensure the orthogonality of 4-velocity and the co-frame, since the latter
is treated as independent variable). The Lagrange multipliers are 0-forms.

Let us derive the Euler-Lagrange equations. To summarise, the independent
variables here are: the metric-affine gravitational field $g_{\alpha\beta}$,
\ $\vartheta^{\alpha}$,\ $\Gamma^{\ \alpha}_{\beta}$, 
the material variables $b_{A},\ b^{B},\ \rho, \mu^{A}_{\ B}$,\ $s,\ X$,
and the Lagrange multipliers $\lambda$.

Varying the action with respect to the latter, one obtains the constraints
(3.1), (3.2), supplemented by $b^{A}\wedge u=0$, and equations (3.3)-(3.5). 
Variations of $s$ and $X$ yield the equations for the pair of Lagrange 
multipliers,
$$
\eta\Big({\partial\varepsilon \over\partial s}\Big)+d(\lambda_{3}u)=0,\ \ \  
d(\lambda_{2}u)=0,\eqno(\z)
$$
while the variations of $\rho$ and $\mu^{A}_{\ B}$ yield respectively
$$
\eta\Big({\partial\varepsilon \over \partial\rho}\Big) -
{1\over 2}\mu^{A}_{\ B}b^{B}_{\alpha}u\wedge Db^{\alpha}_{A} -
u\wedge d\lambda_{1} = 0,\eqno(\z)
$$
$$
\eta\Big({\partial\varepsilon \over \partial\mu^{A}_{\ B}}\Big) =
{1\over 2}\rho b^{B}_{\alpha}u\wedge Db^{\alpha}_{A}.\eqno(\z)
$$
These in fact provide the explicit form of thermodynamical variables:
the pressure and the rotation+deformation tensor, conjugated to hypermomentum.

To finish with the material variables, let us write the equations which result
from the variation of the action (3.6) with respect to $u$, $b^{B}$, $b_{A}$. 
These read respectively: 
$$
-{1\over 2}\rho\mu^{A}_{\ B}b^{B}_{\alpha}Db^{\alpha}_{A} -
\rho d\lambda_{1} + \lambda_{2}dX + \lambda_{3}ds - 2\lambda_{0}\ast u - 
\lambda^{A}\ast b_{A} - \tilde{\lambda}_{A}b^{A}=0,\eqno(\z)
$$
$$
{1\over 2}\rho\mu^{A}_{\ B}\ast(u\wedge D b^{\alpha}_{A})\eta_{\alpha} + 
\lambda^{A}_{B}b_{A} + \tilde{\lambda}_{B}u=0,\eqno(\z)
$$
$$
{1\over 2}\ast D(\rho\mu^{A}_{\ B}b_{\alpha}^{B}u)\vartheta^{\alpha} 
+\lambda^{A}\ast u + \lambda^{A}_{B}b^{B}=0.\eqno(\z)
$$
Hereafter, as usually the dual 3-form of the space-time coframe is denoted
$\eta^{\alpha}:=\ast\vartheta^{\alpha}$. 

It is worth to note, that the 
variations of the matter triad {\it components} look rather non-trivial:
$$
\eta\delta b^{\alpha}_{A}=\vartheta^{\alpha}\wedge\delta b_{A} + 
\delta\vartheta^{\alpha}\wedge b_{A} - b^{\alpha}_{A}({1\over 2}g^{\rho\sigma}
\delta g_{\rho\sigma}\eta +\delta\vartheta^{\beta}\wedge\eta_{\beta}),\eqno(\z)
$$
$$
\eta\delta b^{B}_{\alpha}=\delta b^{B}\wedge\eta_{\alpha} -
\delta\vartheta^{\beta}\wedge\eta_{\alpha}b^{B}_{\beta}.\eqno(\z)
$$

Multiplying (3.10) by $u$, and using (3.1)-(3.5), (3.7), we get the
Lagrange multiplier
$$
2\lambda_{0}=
-{1\over 2}\rho\mu^{A}_{\ B}b^{B}_{\alpha}\ast(u\wedge Db^{\alpha}_{A}) 
-\rho\ast(u\wedge d\lambda_{1}) = 
\rho\Big({\partial\varepsilon \over \partial\rho}\Big).\eqno(\z)
$$
Analogously, the remaining Lagrange multipliers are obtained from the 
exterior products of (3.11) with $\ast u$, (3.11) with $b^{C}$, and (3.12) 
with $u$, respectively,
$$
\tilde{\lambda}_{A}=
-{1\over 2}\rho\mu^{B}_{\ A}u_{\alpha}\ast(u\wedge Db^{\alpha}_{B}),\eqno(\z)
$$
$$
\lambda^{A}_{B}=
{1\over 2}\rho\mu^{C}_{\ B}\ast (u\wedge b_{C}^{\alpha}Db_{\alpha}^{A}),
\eqno(\z)
$$
$$
\lambda^{A}=-{1\over 2}\ast D(\rho\mu^{A}_{\ B}b^{B}_{\alpha}u)u^{\alpha}.
\eqno(\z)
$$

Finally, multiplying (3.12) by $b_{C}$, and using (3.17), we get the
{\it equations of motion} of the specific hypermomentum density:
$$
u\wedge (d\mu^{A}_{\ B} + \mu^{A}_{\ C}b^{\alpha}_{B}Db_{\alpha}^{C} - 
\mu^{C}_{\ B}b^{\alpha}_{C}Db_{\alpha}^{A}) = 0.\eqno(\z)
$$
This generalises the well known equation of motion of spin in the Weysenhoff
model.

Let us find the {\it matter currents}. In the previous section we discussed
these phenomenologically, but now the rigorous derivation from the variational
principle is straightforward. One obtains:
$$
\sigma^{\alpha\beta}:=2{\delta L \over \delta g_{\alpha\beta}}=
\eta [\varepsilon g^{\alpha\beta} + 2\lambda_{0}(u^{\alpha}u^{\beta} - 
g^{\alpha\beta}) + 2\lambda^{A}b^{(\alpha}_{A}u^{\beta)} -
g^{\alpha\beta}\lambda^{A}_{A} - {1\over 2}g^{\alpha\beta}
\ast D(\rho\mu^{A}_{\ B}b^{B}_{\gamma}u)b^{\gamma}_{A} ],\eqno(\z)
$$
$$
\Delta^{\alpha}_{\ \beta}:={\delta L \over \delta \Gamma^{\ \beta}_{\alpha}}=
{1\over 2}\rho\mu^{A}_{\ B}b^{B}_{\beta}b^{\alpha}_{A}u,\eqno(\z)
$$
$$
\Sigma_{\alpha}:={\delta L \over \delta\vartheta^{\alpha}} =
\varepsilon\eta_{\alpha} - 2\lambda_{0}(\eta_{\alpha}-u_{\alpha}u) +
$$
$$
+u_{\alpha}\lambda^{A}b_{A} + g_{\alpha\beta}b^{\beta}_{A}\lambda^{A}u -
\lambda^{A}_{A}\eta_{\alpha} - {1\over 2}\eta_{\alpha}b^{\beta}_{A}\ast 
D(\rho\mu^{A}_{\ B}b^{B}_{\beta}u) + {1\over 2}\eta_{\beta}\ast 
D(\rho\mu^{A}_{\ B}b^{B}_{\alpha}b^{\beta}_{A}u).\eqno(\z)
$$

These expressions are simplified greatly if we denote
$$
J^{\alpha}_{\ \beta}= 
{1\over 2}\rho\mu^{A}_{\ B}b^{\alpha}_{A}b^{B}_{\beta},\eqno(\z)
$$
and introduce the {\it pressure} in a standard way,
$$
p:=\rho\Big({\partial\varepsilon\over \partial\rho}\Big) - 
\varepsilon .\eqno(\z)
$$
We then find 
$$
\lambda_{0}={1\over 2}(\varepsilon + p),\eqno(\z)
$$
and, finally, the matter currents of the hyperfluid read
$$
\sigma^{\alpha\beta}=\eta(\varepsilon u^{\alpha}u^{\beta} - ph^{\alpha\beta})
+ 2u^{\gamma}u^{(\alpha}D\Delta^{\beta)}_{\ \gamma},\eqno(\z)
$$
$$
\Delta^{\alpha}_{\ \beta}=u J^{\alpha}_{\ \beta},\eqno(\z)
$$
$$
\Sigma_{\alpha}= \varepsilon uu_{\alpha} - p(\eta_{\alpha}-uu_{\alpha}) +
2uu^{\beta}g_{\gamma [\alpha }\dot{J}^\gamma {}_{\beta ]}.\eqno(\z)
$$ 
In the derivation of these the equations of the hypermomentum (3.19) 
were used. 

Hypermomentum dynamics can be more conveniently described with respect to an 
anholonomic space-time frame, using the definition (3.23) and the fact that 
the material frame spans the space, orthogonal to the 4-velocity, which is 
expressed by the identity
$$
b^{A}_{\alpha}b^{\beta}_{A}=h^{\beta}_{\alpha}=
\delta^{\beta}_{\alpha}-u^{\beta}u_{\alpha},\eqno(\z)
$$
straightforwardly derivable from (3.1), (3.2). Multiplying (3.19) by 
${1\over 2}\rho b^{\alpha}_{A}b^{B}_{\beta}$, one finds
$$
D\Delta^{\alpha}_{\ \beta} = u^{\alpha}u_{\lambda}D\Delta^{\lambda}_{\ \beta} +
u_{\beta}u^{\lambda}D\Delta^{\alpha}_{\ \lambda}.\eqno(\z)
$$

Comparison of (3.26)-(3.28) with the expressions of matter currents (2.11), 
(2.12), (2.17), (2.16), shows complete agreement of the rigorous 
variational theory with the phenomenological approach. There is, though, one 
refinement --- the hypermomentum density (3.23) is subject to the constraint
$$
J^{\alpha}_{\ \beta}u^{\beta}=J^{\alpha}_{\ \beta}u_{\alpha}=0,\eqno(\z)
$$
which is the analogue of the well known Frenkel condition. Physically this
is motivated by the properties of spin as a part of the total hypermomentum.

A new point is that the variational theory yields the equations of
motion of hypermomentum (3.30), which the phenomenological approach could not
provide in view of the absence of a definition of the phenomenological matter 
currents in terms of the true dynamical variables of the hyperfluid.

\bigskip
\bigskip
\sectio{\bf Conclusion}
\bigskip
\bigskip

The hyperfluid represents a classical model of matter with hypermomentum 
which is the source of the metric-affine gravity. It is interesting to obtain
the exact non-vacuum solutions for the gravitational field equations in 
astrophysical and cosmological (very early stages) setting, this work is now
in progress. Further development of the hyperfluid model should give an 
answer to an important question: is it possible to recover this medium in a
semiclassical treatment of the quantum matter with hypermomentum --- such as,
for example, the manifields [26-28]? 

\bigskip
\bigskip
\centerline{\bf Acknowledgements}
\bigskip
The authors would like to thank Friedrich W. Hehl and Eckehard W. Mielke
for stimulating discussions and useful comments. The research of YNO was
supported by the Alexander von Humboldt Foundation.
\vfill\eject

\bigskip
\bigskip
\centerline{\bf References}
\bigskip
\newref
[1] C. M\o ller, {\it The theory of relativity} (Clarendon Press: Oxford, 
1952).
\newref
[2] J.L. Synge, {\it Relativity -- the general theory} (North-Holland: 
Amsterdam, 1960).
\newref
[3] J. Ehlers, in: {\it Recent developments in general relativity}, Festschrift
for L.Infeld (Pergamon Press/PWN: Oxford/Warsaw, 1962) 201.
\newref
[4] J. Weyssenhoff, and A. Raabe, {\sl Acta Phys. Polon.} {\bf 9} (1947) 7.
\newref
[5] F. Halbwachs, {\it Th\'eorie relativiste des fluides \`a spin} 
(Gauthier-Villars: Paris, 1960).
\newref
[6] I. Bailey, and W. Israel, {\sl Commun. Math. Phys.} {\bf 42} (1975) 65.
\newref
[7] I. Bailey, {\sl Ann. Phys. (USA)} {\bf 119} (1979) 76.
\newref
[8] W. Kopczy\'nski, {\sl Phys. Rev.} {\bf D34} (1986) 352.
\newref
[9] F.W. Hehl, P. von der Heyde, G.D. Kerlick, and J.M. Nester, {\sl Revs.
Mod. Phys.} {\bf 48} (1976) 393.
\newref
[10] A. Trautman, {\sl Symposia Mathematica} {\bf 12} (1973) 139.
\newref
[11] F.W. Hehl, G.D. Kerlick, and P. von der Heyde, {\sl Z. Naturforsch.}
{\bf 31a} (1976) 111; 524; 823; and {\sl Phys. Lett.} {\bf B63} (1976) 446.
\newref
[12] L.L. Smalley, {\sl Phys. Lett.} {\bf A61} (1977) 436.
\newref
[13] F.W. Hehl, and G.D. Kerlick, {\sl Gen. Rel. Grav.} {\bf 9} (1978) 691.
\newref
[14] F.W. Hehl, E.A. Lord, and L.L. Smalley, {\sl Gen. Rel. Grav.} {\bf 13}
(1981) 1037.
\newref
[15] F.W. Hehl, E.A. Lord, and Y. Ne'eman, {\sl Phys. Lett.} {\bf B71} (1977)
432; and {\sl Phys. Rev.} {\bf D17} (1978) 428.
\newref
[16] V.N. Ponomariev, and Yu.N. Obukhov, {\sl Gen. Rel. Grav.} {\bf 14} (1982)
309.
\newref
[17] E.A. Lord, {\sl Phys. Lett.} {\bf A65} (1978) 1.
\newref
[18] F.W. Hehl, J.D. McCrea, and E.W. Mielke, {\sl Found. Phys.} {\bf 19} 
(1989) 1075.
\newref
[19] F.W. Hehl, J.D. McCrea, and E.W. Mielke, in: {\sl Exact Sciences and their
Philosophical Foundations}. {\it Vortr\"age des Internationalen 
Hermann-Weyl-Kongresses, Kiel 1985} /Eds. W. Deppert, K. H\"ubner, 
A. Oberschelp, and V. Weidemann (P.Lang Verlag: Frankfurt, 1988) 241;
R.D. Hecht, and F.W. Hehl, {\sl Proc. 9th Italian Conf. Gen. Rel. and Grav.
Phys., Capri (Napoli)}. /Eds. R. Cianchi et al (World Scientific: Singapore, 
1991) p. 246.
\newref
[20] A.Trautman, in: {\sl General Relativity and Gravitation}. {\it One Hundred
Years After the Birth of Albert Einstein} /Ed. A. Held (Plenum Press: New York,
1980), vol. 1, p. 287. 
\newref
[21] Yu.N. Obukhov, and V.A. Korotky, {\sl Class. Quantum Grav.} {\bf 4} (1987)
1633.
\newref
[22] E. Cosserat, and F. Cosserat, {\it Th\'eorie des corps d\'eformables} 
(A.Hermann et Fils: Paris, 1909).
\newref
[23] R.D. Mindlin, {\sl Arch. Rat. Mech. Anal.} {\bf 16} (1964) 51.
\newref
[24] J.R. Ray, and L.L. Smalley, {\sl Phys. Rev. Lett.} {\bf 49} (1982) 1059.
\newref
[25] R. De Ritis, M. Lavorgna, G. Platania, and C. Stornaiolo, {\sl Phys. Rev.}
{\bf D28} (1983) 713.
\newref
[26] Y. Ne'eman, {\sl Proc. Nat. Acad. Sci. (USA)} {\bf 74} (1977) 4157.
\newref
[27] Y. Ne'eman, and Dj. \v{S}ija\v{c}ki, {\sl Ann. Phys. (NY)} {\bf 120} 
(1979) 292.
\newref
[28] Y. Ne'eman, and Dj. \v{S}ija\v{c}ki, {\sl Phys. Lett. B157} (1985) 275;
and {\sl Phys. Lett.} {\bf B200} (1988) 489.

\end